\documentclass[showkeys,eqsecnum,twocolumn,showpacs,preprintnumbers,amssymb,aps]{revtex4}
\usepackage{graphicx}
\usepackage{slashbox}
\usepackage{dcolumn}
\usepackage{bm}
\usepackage{latexsym,epsfig}

\begin{document}

\preprint{\today}

\title{Studies of all order correlation effects in the isotope shifts and fine structure variation constants: Applications to Na and Mg$^+$}
\vspace{0.5cm}

\author{B. K. Sahoo \footnote{Email: bijaya@prl.res.in}}
\affiliation{Theoretical Physics Divison, Physical Research Laboratory, Ahmedabad-380009, India}

\date{\today}
\vskip1.0cm

\begin{abstract}
We study the electron correlation effects in the calculations of isotope shifts in Na and Mg$^+$ using 
the relativistic coupled-cluster method. The trends of the correlation effects
are explicitly discussed and comparison of the present results with the previously
reported results are given. We also present the fine structure constant variation 
coefficients for many states in the above systems. From these results, it
is possible to find out suitable transition lines those can be used as
anchor and probe lines
in finding possible variation in the fine structure constant.
\end{abstract}


\maketitle

\section{Introduction}
Study of isotope shifts (ISs) in atomic systems has been a long time immense
interest to the physicists \cite{breit,king, otten1, theodossiou}. It is
a challenging problem to find out the role of the correlation effects in the
IS calculations using
an {\it ab initio} approach for a system involving more than three electrons.
The most difficult part in IS determination lies in calculating the specific 
mass shift (SMS) which involves two-body interactions between the electron 
momenta \cite{king,breit} and treating correlation effects to all orders in 
this property is as difficult as considering the Coulomb interaction between the electrons.
Often semi-emperical methods \cite{otten1,otten2} have been used to 
calculate these properties without describing the electron correlation 
behaviors. There have been a number of works carried out using the
non-relativistic theories to include the correlation effects that consider
particular type of correlation effects to all orders and some of the correlation
effects are taken only up to finite order \cite{martensson, lindroth,pendrill,kuth,fischer}. 

Studies of ISs are important in a number of applications ranging
from nuclear physics to astrophysics \cite{otten1,otten2,chen,wendt,kozlov,brostroem,van}. Accurate determination of 
these quantities in combinations with the corresponding measurements can
be used to extract the relative root mean square (rms) nuclear charge radii
of different isotopes that cannot be measured directly \cite{otten1,otten2}. They are 
also used to test various nuclear models \cite{chen,wendt}. Since the rms radii
variation in the considered systems can be determined precisely using the
nuclear models due to small sizes of their nuclei, we will not focus our
studies in that direction here. Again, some of the ISs
are also used to probe the isotopic abundances of different elements
in the astronomical objects \cite{brostroem,van}. 

In another context, it is believed that there is possibility of space-time 
variation of the fine-structure constant $\alpha=e^2/hc$ at the
cosmological scale \cite{bronnikov, webb, chako}. This can be verified by
analyzing the quasar absorption spectra \cite{srianand,chand,tzanavaris,quast}. 
It is also possible to find out any variation of $\alpha$ by comparing the
measurements of the ratios of the fine structure intervals to the optical 
transition frequencies in the laboratory experiments and that from some distant 
astrophysical objects \cite{dzuba1,berengut1,dzuba2}. Any difference in 
these results can be related to the space-time variation of $\alpha$.
However, the frequency shifts that corresponds to such transitions are
of the same order of magnitude with the possible isotope shifts in the
elements present in the astrophysical objects \cite{kozlov}. Any changes in the isotope abundances
in these objects will be the systematic errors in the finding of $\alpha$
variation \cite{levshakov, varshalovich, korol}. As suggested by Kozlov 
{\it et al} \cite{kozlov} it is possible to construct anchor lines,
those are insensitive to the variation of abundances in the astrophysical 
objects, and probe lines, those are insensitive to the $\alpha$ variation.
It is also possible to construct such lines by considering combinations 
of suitable transitions of the elements present inside the astrophysical
objects \cite{kozlov}. To find out the sensitive lines to $\alpha$ variation
and/or strong IS in any system, it is necessary to determine the $\alpha$ variation
coefficient $q$ and IS parameters, respectively, as defined explicitly
below. Since both Na and Mg$^+$ are proposed as suitable candidates for this 
purpose \cite{berengut1,dzuba2}, we would like to determine these properties
accurately in the above systems. 

There are three relativistic many-body methods have already been used to 
calculate the IS constants in the considered systems \cite{safronova,berengut2,korol}. Safronova and Johnson
have used at most two orders of Coulomb interactions through the third order
many-body perturbation theory (MBPT(3) method) to calculate the ISs and the
trends of different electron correlation effects are presented at this level of
approximation \cite{safronova}. Berengut {\it et al.} \cite{berengut2} have
considered the IS operators in the atomic Hamiltonian with a varying parameter, 
then they calculate the energies self-consistently. Changes in the energies are
estimated to the first order in that parameter to extract the corresponding ISs.
 However, it is not possible to find out the role of different electron
correlation effects through this procedure. The individul field shift (FS) and 
specific mass shift (SMS) results reported by these two methods do not match 
well with each other although the final ISs reported by both the works are 
close to the experimental measurements. This is due to the fact that there is a
large cancellation between the initial state and final state results. However,
it does not really 
explain the reason of discrepancies at the individual result. Korol and
Kozlov \cite{korol} have followed approach considered by Dzuba {\it et al.}
and present only the final results without giving details of the correlation
contributions. Therefore, it is necessary to employ a relativistic many-body
method that accounts correlation effects to all orders and it enables us to
understand the underlying role of different correlation effects in these 
calculations. To fill-up the gap between the differences in the individual
results reported by Safronova and Johnson and Berengut {\it et al.}, we 
follow an approach similar to Safronova and Johnson but consider atomic
wavefunctions determined using the relativistic coupled-cluster (RCC) theory
which is an all order perturbation method having size consistent property
\cite{lindgren,szabo}. The catagorily distinct features of this work are:
(i) it will explain the differences between the results reported in 
\cite{safronova} and \cite{berengut2}, (ii) it will present individual 
contributions from various correlation (RCC) terms to both FS and SMS
constant calculations and (iii) it will give $q$ and ISs of many low-lying 
transitions in the considered systems those are not studied in the literature to date.

 The remaining part of this paper is arranged as follwing: In the next section
we will give the general theory of fine structure constant variation and ISs
in the atomic systems. In the following section, we will explain the many-body 
techniques in the RCC framework that has been used to calculate the energies
and IS properties. In Sec. IV, we will present the results and discuss
them by comparing with the earlier results. In the last section, we will summarise
our work and draw the conclusions.

\section{Theory}
The IS ($\delta E_{v}^{IS}$) to an energy level of state $|\Psi_v\rangle$ in an atomic system is
mainly classified into two parts: (a) the shifts ($\delta E_{v}^{MS}$) due to 
the consideration of finite nuclear mass that is known as mass shift (MS) and
(b) the shifts ($\delta E_{v}^{FS}$) caused due to change in the nuclear 
charge radius from one isotope to another, known as FS. The MS can be further
divided into normal mass-shift (NMS) and SMS. The NMS is given by
\begin{eqnarray}
\delta E_{v}^{NMS} &=& - \frac {m_e} {m_e+M_A} E_v^{theory} \nonumber \\
                    &=& - \frac {m_e} {M_A} E_v^{expt},
\label{eqn1}
\end{eqnarray}
where $m_e$ and $M_A$ are the electron and atomic masses, respectively. Here
$E_v^{theory}$ and $E_v^{expt}$ are the respective energy levels of state $v$ 
that can be theoretically calculated and experimentally observed, respectively.
Accuracy in $E_v^{theory}$ depends on the theoretical method employed in the 
calculations
while accuracy of $E_v^{expt}$ depends on the measurement techniques. For the
considered systems, we use $E_v^{expt}$ to estimate the corresponding NMSs.

The specific mass shift for the state $|\Psi_v\rangle$ is given by
\begin{eqnarray}
\delta E_{v}^{SMS} &=& - \frac {M_A} {2(m_e+M_A)^2} \langle \Psi_v | \sum_{ i \ge j} \vec p_i \cdot \vec p_j | \Psi_v \rangle \nonumber \\ 
              &=& - \frac {M_A} {(m_e+M_A)^2} \langle \Psi_v | Z | \Psi_v \rangle \nonumber \\
              & \simeq & - \frac {1} {M_A} Z_v ,
\label{eqn2}
\end{eqnarray}
where $\vec p_i$ and $\vec p_j$ and are the momenta of the electrons $i$ and 
$j$, respectively. For the calculation point of view, we define $ Z$ as a
two-body operator and express them in the second quantized notation as
\begin{eqnarray}
Z= \frac {1}{2} \sum_{ab,cd} z_{abcd} a_a^{\dagger} a_b^{\dagger} a_d a_c,
\label{eqn3}
\end{eqnarray}
with
\begin{eqnarray}
 z_{abcd} &=& \langle ab | \vec p_i \cdot \vec p_j | cd \rangle.
\label{eqn4}
\end{eqnarray}

In the angular matrix form, we can have
\begin{eqnarray}
 z_{abcd} &=& (-1)^{j_a-m_a+j_b-m_b} X(abcd) \sum_{q=-1,0,1} (-1)^q \nonumber \\
   && \left ( 
         \matrix {
         j_a & 1 & j_c \cr
         -m_a & q & m_c \cr
                 }
         \right ) 
  \left (
         \matrix {
         j_b & 1 & j_d \cr
         -m_b & -q & m_d \cr
                 }
         \right ), 
\label{eqn5}
\end{eqnarray}
where the reduced matrix element is given by
\begin{eqnarray}
X(abcd) &=& \sqrt{(2j_a+1)(2j_b+1)(2j_c+1)(2j_d+1)} \nonumber \\
  &&  \left (
         \matrix {
         j_a & 1 & j_c \cr
         1/2 & 0 & -1/2 \cr
                 }
         \right ) 
   \left (
         \matrix {
         j_b & 1 & j_d \cr
         1/2 & 0 & -1/2 \cr
                 }
         \right ) \nonumber \\ && (-1)^{j_a+j_b+1} R(ac) R(bd). 
\label{eqn6}
\end{eqnarray}

Using the relation $|\vec p_i|=p_i= \frac{d}{dr_i} - \frac {\kappa_i}{r_i}$, 
we can have the expressions for the radial functions as
\begin{widetext}
\begin{eqnarray}
 R(ij) &=& -i \int_0^{\infty} dr \left [ P_i \left ( \frac{dP_j}{dr_j} + \frac {\vartheta(\kappa_i,\kappa_j)P_j}{r_j} \right ) + Q_i \left ( \frac{dQ_j}{dr_j} + \frac {\vartheta(-\kappa_i,-\kappa_j)Q_j}{r_j} \right ) \right ],
\label{eqn7}
\end{eqnarray}
\end{widetext}
where $P_i$ and $Q_i$ are the large and small components of the Dirac orbital 
of electron $i$ and the expression is valid only for
\begin{eqnarray}
\vartheta(\kappa_i,\kappa_j) = \left \{ 
\begin{array}{l l}
  -\kappa_j & \text{for $\kappa_i = \kappa_j-1$} \\
  -\kappa_j & \text{for $\kappa_i = -\kappa_j$} \\
  \kappa_j+1 & \text{for $\kappa_i = \kappa_j+1$} \\
\end{array} \right.
\label{eqn8}
\end{eqnarray}
otherwise its value is zero.

The above expression can also be given in another form by using the relation 
$p_i= m_e c |\vec \alpha|$ with $c$ is the velocity of light and $\vec \alpha$ 
being the Dirac matrix as
\begin{widetext}
\begin{eqnarray}
 R(ij) &=& -i m_e c \int_0^{\infty} dr \left [ \left ( \kappa_i -\kappa_j -1 \right ) P_i Q_j +  \left ( \kappa_i -\kappa_j +1 \right ) Q_i P_j  \right ].
\label{eqn9}
\end{eqnarray}
\end{widetext}

Since the radial functions involved in both the expressions given by 
Eqs. (\ref{eqn8}) and (\ref{eqn9}) are different, they can be simultaneously
used by a given many-body method to verify the numerical inaccuracies
in the calculations. 

For the field-shift calculation, we consider the nucleus is uniformly charged
sphere of radius R with Z number of protons so that an electron in a distance 
r sees the nuclear potential
\begin{eqnarray}
V_{nuc}(r,R) = \left\{ 
\begin{array}{l l}
  -(Z/2R) [3- r^2/R^2] & \text{for $r < R$} \\
  -Z/r & \text{for $r \ge R$} \\
\end{array} \right.
\label{eqn10}
\end{eqnarray}
This assumtion will give reasonably accurate results for the considered light 
systems. In this case, the rms $\langle r^2 \rangle$ is given by
\begin{eqnarray}
\langle r^2 \rangle = \frac{3}{5} R^2.
\label{eqn11}
\end{eqnarray}
 
Now any change $\delta \langle r^2 \rangle^{A,A'}$ in $\langle r^2 \rangle$ for 
different isotopes $A$ and $A'$ cause FS and is given by
\begin{eqnarray}
\delta E_{IS}^{FS} &=& - F \lambda^{A,A'}
\label{eqn12}
\end{eqnarray}
where $F$ is called as field shift constant and $ \lambda^{A,A'}$ is an
expansion of $\delta \langle r^2 \rangle$ as
\begin{eqnarray}
\lambda^{A,A'} = \sum_n \frac{C_n}{C_1} \delta \langle r^2 \rangle^{A,A'}
\label{eqn13}
\end{eqnarray}
for $C_n$ being the expansion coefficients which are given by Seltzer \cite{seltzer}. In the present 
studies, it is sufficient enough to neglect the higher terms and assume
$ \lambda^{A,A'} \simeq \delta \langle r^2 \rangle^{A,A'}$.

It can be shown from the above discussions that
\begin{eqnarray}
F &=& \langle \frac{\delta V_{nuc}(r,R)}{\delta \langle r^2 \rangle}  \rangle \nonumber \\
  &=& \left\{ 
\begin{array}{l l}
  -(5Z/4R^2) [1- r^2/R^2] & \text{for $r \le R$} \\
  0 & \text{for $r > R$} \\
\end{array} \right.
\label{eqn14}
\end{eqnarray}
The above operator is a one-body scalar operator and its expectation value
with respect to $|\Psi_v\rangle$ will give its FS constant, $F_v$.

Finally, the IS of any transition between state $i$ to state $f$ of an isotope with 
mass number $M_{A'}$ to another isotope with mass number $M_A$ is given by
\begin{eqnarray}
\delta \nu_{if}^{A',A} &=& \left ( \frac{1}{M_{A'}} - \frac{1}{M_A} \right ) \left [ (E_i^{expt} - E_f^{expt}) + (Z_i - Z_f) \right ] \nonumber \\ &&+ (F_i - F_f) \delta \langle r^2 \rangle^{A',A} \nonumber \\
 &=& \frac{M_A-M_{A'}}{M_{A'}M_A}( k_{if}^{NMS} + k_{if}^{SMS} ) + F_{if} \delta \langle r^2 \rangle^{A',A} .
\label{eqn14a}
\end{eqnarray}

 In the literature, $k_{if}^{NMS}=E_i^{expt} - E_f^{expt}$, $k_{if}^{SMS}=Z_i-Z_f$ and $F_{if}=F_i - F_f$ are known as NMS constant, SMS constant and FS 
constant for the transition $i$ to $f$, respectively.

In the context of $\alpha$ variation, the transition frequency observed from 
some distant astrophysical objects
can be expanded in a series of $\alpha^2$ at $\alpha=0$ in the following way
\begin{eqnarray}
\omega(x) &=& \omega^{(0)} + \alpha^2 \omega^{(0)} + \mathfrak{O}(\alpha^4).
\label{eqn15}
\end{eqnarray}
For a fine-structure transition, the first coefficient is zero while it is
finite for an optical transition. Therefore, the ratios of these transition 
frequencies can have different leading orders in $\alpha^2$. In an
equivalent form, $\omega(x)$ can be expanded at $\alpha=\alpha_0$ for the
laboratory value $\alpha_0$ as
\begin{eqnarray}
 \omega(x) &=& \omega_{lab} + q x +  \mathfrak{O}(x^2)
\label{eqn16}
\end{eqnarray}
where $\omega_{lab}$ is the transition frequency in the present laboratory 
value and $q$ is the $\alpha$ variation coefficient defined according to
$x=(\alpha/\alpha_0)^2 -1$. In principle, $q$ can be determined from the first
derivative $\frac{d\omega(x)}{dx}$ for small value of $x$ and from a numerical 
calculation it can be assumed as
\begin{eqnarray}
 q \approx \frac {\omega(+ \delta x) - \omega( -\delta x)}{2 \delta x}. 
\label{eqn17}
\end{eqnarray}
For simpilicity, if we assume $\delta x=0.05$ then, we have
\begin{eqnarray}
 q \approx 10 [\omega(+ 0.05) - \omega( -0.05)].
\label{eqn18}
\end{eqnarray}

\section{Method of calculations}
\subsection{Wavefunction and energy calculations}
For our purpose, we consider the Dirac-Coulomb Hamiltonian given by
\begin{eqnarray}
H^{DC} = \sum_i \left [ c \vec \alpha \cdot \vec p_i + (\beta -1) c^2 + V_{nuc}(r_i) \right ] + \sum_{i>j} \frac {1}{r_{ij}}, \ \ \
\label{eqn19}
\end{eqnarray}
where we use atomic unit and consider the laboratory value of fine structure 
constant as $c=1/\alpha_0=137.03599972$ in the present calculation. For 
$\alpha$ variation study, we consider different values of $\alpha$ in the above
expression.

Since both Na and Mg$^+$ are one valence configuration systems, we
construct their atomic wavefunctions in the RCC framework as
\begin{eqnarray}
|\Psi_v \rangle = e^T \{ 1+ S_v\} |\Phi_v\rangle, 
\label{eqn20}
\end{eqnarray}
where $|\Phi_v\rangle$ is the Dirac-Fock (DF) wavefunction that is constructed 
by appending a valence electron $v$ to the DF wavefunction of the closed-shell
configuration $[2p^6]$ represented by $|\Phi_0\rangle$; i.e. $|\Phi_v\rangle=a_v^{\dagger} |\Phi_0\rangle$. Here $T$ and $S_v$ are the RCC operators that 
excite electrons from $|\Phi_0\rangle$ and $|\Phi_v\rangle$, respectively, due 
to the electron correlation effects that have been neglected in the DF method. 
The curly bracket with $S_v$ operator represents it is in normal order form. 
Since both the considered systems are small in size, we approximate our RCC
theory only for all possible single and double excitations by defining
\begin{eqnarray}
T &=& T_1 + T_2 \nonumber \\
\text{and} \ \ \ S_v &=& S_{1v} + S_{2v} ,
\label{eqn21}
\end{eqnarray}
which is known as CCSD method.

The detailed equations to solve both $T$ and $S_v$ operator amplitudes are 
discussed in \cite{sahoo,mukherjee}. The electron attachment energy or 
negative of the ionization potential (IP) is evaluated by
\begin{eqnarray}
\Delta E_v = \langle \Phi_v | \overline{H}_N^{DC} \{ 1+ S_v\} |\Phi_v\rangle, 
\label{eqn21}
\end{eqnarray}
where $\overline{H}^{DC}=e^{-T} H^{DC} e^T$ and the normal order Hamiltonian 
is given by $H_N^{DC}= H^{DC} - \langle \Phi_0 | H^{DC} |\Phi_0\rangle$. 
To improve the quality of the results, we construct the triple excitation 
RCC operator for $S_v$ as
\begin{eqnarray}
S_{3v}^{pert} &=& \frac{ \widehat{H_N^{DC} T_2} + \widehat{H_N^{DC} S_{2v}}}{\epsilon_v + \epsilon_a + \epsilon_b - \epsilon_p - \epsilon_q - \epsilon_r},
\label{eqn22}
\end{eqnarray}
and include its effects in Eq. (\ref{eqn21}) to evaluate energy which also
enters into the CCSD amplitude determining equation for $S_v$. This approach
is called as CCSD(T) method and it includes the most important part of the 
triple excitation effects. Here widehat symbol means the
terms are connected and $\epsilon$'s are the single particle energies of 
the orbitals ($+$ sign means for incoming orbitals and $-$ sign means for out 
going orbitals) involved in the excitaions.

\begin{figure}[h]
\includegraphics[width=8.5cm,clip=true]{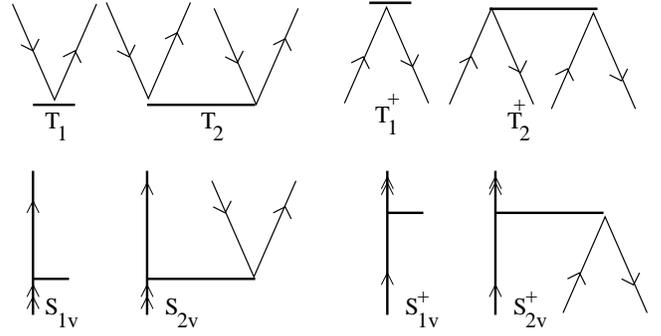}
\caption{Diagrammatic representation of the RCC operators along with their
conjugates. Lines with single arrows going down represent the occupied (hole)
orbitals, lines with single arrows going up represent the virtual (particle)
orbitals and lines with double arrows represent the valence orbitals, 
respectively. Note that a particle line can be converted to a valence 
orbital in a special case, but not the other way around.}
\label{fig1}
\end{figure}

\subsection{Properties evaluation}
 The expectation value of any physical operator $O$ using the RCC method 
for a given state $|\Psi_v \rangle$ is evaluated by
\begin{eqnarray}
\langle O \rangle_v &=& \frac{\langle \Phi_v | \{ 1+ S_v^{\dagger}\} e^{T^{\dagger}} O_N e^T \{ 1+ S_v\} |\Phi_v\rangle}{\langle \Phi_v | \{ 1+ S_v^{\dagger}\} e^{T^{\dagger}} e^T \{ 1+ S_v\} |\Phi_v\rangle},
\label{eqn23}
\end{eqnarray}
where $O_N = O - \langle \Phi_0 | O |\Phi_0\rangle$ is the normal order 
form of $O$. We note that since both $T$ and $S_v$ are in normal order form and 
wavefunctions are calculated using $H_N^{DC}$, so it is necessary to consider
$O_N$ in the above expression; which will be necessary to note for the following discussions.

\begin{figure}[h]
\includegraphics[width=7.0cm,clip=true]{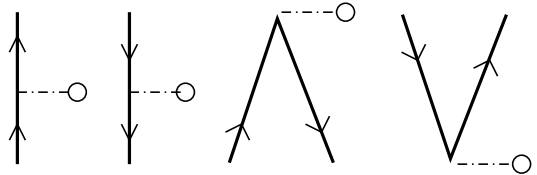}
\caption{Diagrammatic representations of any one-body operator "$O$" as per
the definitions of lines used in Fig. \ref{fig1}.}
\label{fig2}
\end{figure}
 Using the above method, there are a number of studies carried out on varities 
of properties and the procedure to evaluate or approximate the non-truncative 
series like $e^{T^{\dagger}} O_N e^T$ and $e^{T^{\dagger}} e^T$ for the 
required accuracy of the results are given elsewhere (for example see
\cite{sahoo,mukherjee}). However, these calculations were only for the 
properties
involving one-body operators. For the two-body SMS operator $Z$, the normal
order form will have two parts $Z_N = Z_1 + Z_2$ which can be written as
\begin{eqnarray}
Z_1 &=& \sum_{i,j} z_{ij} \{ a_i^{\dagger} a_j \} \nonumber \\
\text{and} \ \ \ Z_2 &=& \frac{1}{2} \sum_{ij,kl} z_{ijkl} \{ a_i^{\dagger} a_j^{\dagger} a_l a_k \},
\label{eqn24}
\end{eqnarray}
where the curly bracket means they are in normal order forms and the amplitudes of the effective one-body operator $Z_1$ is given by
\begin{eqnarray}
 \langle  Z_1  \rangle_{ij} = z_{ij}  &=& \sum_c \left [ \langle ic| Z | jc \rangle - \langle ic| Z | cj \rangle \right ].
\label{eqn25}
\end{eqnarray}
In the above expressions $i,j,k$ and $l$ indices used for any generic orbitals
whereas $c$ represents all the occupied orbitals. The DF result of this 
operator is determined by $Z_1$ by fixing its amplitude for $i=j=v$; for 
a given valence orbital $v$.

\begin{figure}[h]
\includegraphics[width=8.5cm,clip=true]{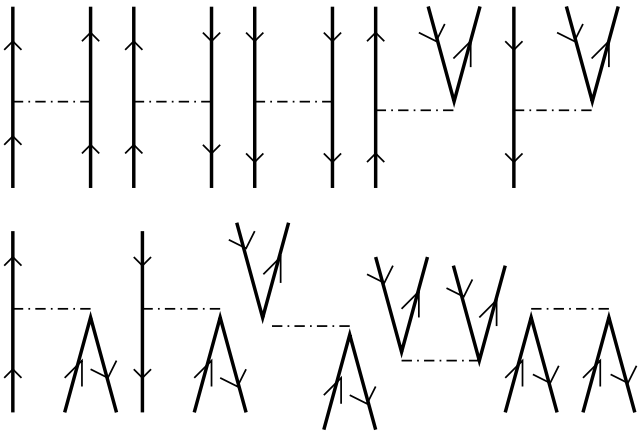}
\caption{Diagrammatic representation of $Z_2$ operator. Electron-electron 
Coulomb interaction ($V_c$) operator can be represented in the same way. To 
differentiate the notation between these two operators, we will use the 
dashed ($- - - - -$) lines for $V_c$ whenever it is necessary.}
\label{fig3}
\end{figure}
 To evaluate contributions from $\overline{Z}_1=e^{T^{\dagger}} Z_1 e^T$,
we have considered the same procedure that is employed for the one-body
operators \cite{sahoo, mukherjee} as discussed above. We also follow a
similar proedure to account contributions from $\overline{Z}_2=e^{T^{\dagger}} Z_2 e^T$. First we evaluate the effective one-body terms from $\overline{Z}_2$ 
at the same level of approximations of $\overline{Z}_1$ and consider along 
with it. It can be noted that the effective two and three body terms 
from $\overline{Z}_1$ considered earlier \cite{sahoo, mukherjee} are of 
at most forth order in Coulomb interaction; otherwise they were neglected. In
the present work, we also assume same approximation to evaluate the effective 
two and three body terms from $\overline{Z}_2$; however the number of Goldstone
diagrams in this case increase drastically compared to the earlier case. We
discuss below the crucial contributing effective one-body and two-body
diagrams explicitly.

\begin{table}[h]
\caption{Different parameters (parm) used for the considered symmetries ($l$) 
for the wavefunction calculations in Na and Mg$^+$. Here $N^{GTO}$ and $N^{CC}$
are the total number of GTOs used to generate the single particle orbitals
and considered active orbitals for the RCC calculations, respectively.}
\begin{tabular}{c|ccccc}
\hline
\hline
\backslashbox{parm}{$l$} & 0  & 1 & 2 & 3 & 4 \\
\hline
 & & & \\
$\eta_0 \times 10^{2}$ & 0.450 & 0.452 & 0.460 & 0.480 & 0.510 \\
$\zeta$ & 2.69 & 2.69 & 2.69 & 2.69 & 2.73 \\
$N^{GTO}$ & 34 & 32 & 30 & 30 & 30 \\
$N^{CC}$ & 13 & 13 & 12 & 11 & 9 \\
\hline
\hline
\end{tabular}
\label{tab1}
\end{table}
\subsection{Goldstone diagrammatic representation}
To understand the role of different correlation effects coming through 
various RCC terms, we express $T$ and $S_v$ operators diagrammatically 
in Fig. \ref{fig1} with their corresponding conjugates. It has to be 
noted here that the core correlation effects are taken care through the
$T$ and its conjugate operators. The lower order pair-correlation effects
and core-polarization effects involving the valence orbitals are accounted
through $S_{1v}$ and $S_{2v}$ operators, respectively (for example see 
\cite{sahoo} and references therein). 

There are only four general Goldstone diagrams representing any normal ordered 
one-body operator which are shown in Fig. \ref{fig2}. Therefore, the 
contributions from the FS operator $F$ and one-body part of the SMS operator
$Z_1$ will be described by these diagrams.

We also give the diagrammatric representation of the $Z_2$ operator in
Fig. \ref{fig3} and
the Coulomb operator $V_c=\frac {1}{r_{ij}}$ can also be represented by 
similar diagrams. 

Now using the above diagrammatic tools, we will be able to discuss the roles
of various correlation effects.

\begin{table}[h]
\caption{Calculated IPs (in $cm^{-1}$) for different states in Na with different values of $\alpha$.}
\begin{center}
\begin{tabular}{lrrr}
\hline
\hline
 Atomic & IP with & IP with & IP with \\
 state & $\omega(0.0)$ & $\omega(+0.05)$ & $\omega(-0.05)$ \\
\hline
  &  &  \\
$3s \ ^2S_{1/2}$ & 41345.396 & 41348.311 & 41342.485 \\
$3p \ ^2P_{1/2}$ & 24461.150 & 24461.792 & 24460.506 \\
$3p \ ^2P_{3/2}$ & 24442.953 & 24442.677 & 24443.230 \\
$4s \ ^2S_{1/2}$ & 15676.945 & 15677.627 & 15676.262 \\
$3d \ ^2D_{5/2}$ & 12269.785 & 12269.785 & 12269.784 \\
$3d \ ^2D_{3/2}$ & 12269.738 & 12269.736 & 12269.740 \\
$4p \ ^2P_{1/2}$ & 11148.860 & 11149.072 & 11148.648 \\
$4p \ ^2P_{3/2}$ & 11142.482 & 11142.393 & 11142.571 \\
$5s \ ^2S_{1/2}$ &  7708.343 &  7708.789 &  7707.896 \\
$4d \ ^2D_{5/2}$ &  6781.742 &  6781.743 &  6781.742 \\
$4d \ ^2D_{3/2}$ &  6781.702 &  6781.701 &  6781.703 \\
$4f \ ^2F_{5/2}$ &  6847.330 &  6847.331 &  6847.329 \\
$4f \ ^2F_{7/2}$ &  6847.323 &  6847.323 &  6847.322 \\
$5p \ ^2P_{1/2}$ &  4930.127 &  4930.414 &  4929.841 \\
$5p \ ^2P_{3/2}$ &  5117.936 &  5117.826 &  5118.047 \\
\hline
\hline
\end{tabular}
\end{center}
\label{tab2}
\end{table}

\section{Results and Discussions}

\subsection{Single particle orbitals}
In the present calculations, we have used Gaussian type orbitals (GTOs) 
basis to construct the single particle orbitals which are given by
\begin{eqnarray}
F_j^{GTO}(r_i) = r_i^{n_{\kappa}} e^{-\eta_j r_i^2}
\label{eqn26}
\end{eqnarray}
where $\eta_j$ is an arbitrary parameter which is again defined by two
parameters as
\begin{eqnarray}
\eta_j = \eta_0 \zeta^{j-1}
\label{eqn27}
\end{eqnarray}
for $j$ representing the number of GTOs ($F_j^{GTO}(r_i)$) considered for a 
given symmetry with the radial quantum number $n_{\kappa}$. The radial
grid points to define these functions are generated using the relation
\begin{eqnarray}
r_i = r_0 \left [ e^{h(i-1)} - 1 \right ]
\label{eqn27}
\end{eqnarray}
where the starting grid point and the step size are considered as 
$r_0= 2 \times 10^{-6}$ and $h=0.03$, respectively, for a total number
of grid points, reprsented by $i$, 750.

The above orbitals are also geneted by accounting the finite size of the 
nucleus assuming a two-parameter Fermi-nuclear-charge
distribution that is given by
\begin{equation}
\rho(r_i) = \frac {\rho_0} {1 + e^{(r_i-c)/a}},
\label{eqn29} 
\end{equation} 
where $\rho_0$ is the density for the point nuclei, $c$ and $a$ are the
half-charge radius and skin thickness of the
nucleus. These parameters are chosen as
\begin{equation}
a = 2.3/4(ln3)
\label{eqn30}
\end{equation}
and
\begin{equation}
c = \sqrt{ \frac{5}{3} r_{rms}^2 - \frac{7}{3} a^2 \pi^2},
\label{eqn31}
\end{equation}
where $r_{rms}$ is the root mean square radius of the corresponding nuclei
which is determined by
\begin{equation}
r_{rms} = 0.836 M_A^{1/3} + 0.570
\label{eqn32}
\end{equation}
in $fm$ for the atomic mass $M_A$. It should be noted that $r_{rms}$ can also be
obtained by combining the measured ISs with our calculated quantities for FS 
and MS, but we use the above approximation to calculate the final IS of
many transitions where the experimental results are not available. Again
both Na and Mg$^+$ have lighter nuclei, so the above approximation gives
reasonably accurate $r_{rms}$ values for the considered systems. 

We have considered same $\eta_0$, $\zeta$ and GTO parameters for different
symmetries in both Na and Mg$^+$ in order to compare the trends of the
correlation effects due to their nuclear sizes. Although a large 
number of GTOs are used to produce the single particle orbitals, 
the higher energy orbitals do not contribute significantly in the RCC
calculations but they increase the amount of computations. Therefore, we
have considered only the required active orbitals along with all the
occupied orbitals in the RCC calculations those play significant roles in
the accurate determination of the considered properties. A brief summary of
the considered parameters that are taken in the DF and RCC calculations are
listed in Table \ref{tab1}.

\begin{table*}[h]
\caption{Excitation energies and $q$ values in $cm^{-1}$ for different states in Na.}
\begin{center}
\begin{tabular}{lccccc}
\hline
\hline
 Transition & Experiment & \multicolumn{2}{c}{\underline{Others \cite{berengut1}}} & \multicolumn{2}{c}{\underline{This work}} \\
  &  EE \cite{nist} & EE & $q$ & EE & $q$ \\
\hline
  &  &  \\
$3s \ ^2S_{1/2} \rightarrow 3p \ ^2P_{1/2}$ & 16956.172 & 16858 & 45(4) & 16884.246 & 45.4(2) \\
$3s \ ^2S_{1/2} \rightarrow 3p \ ^2P_{3/2}$ & 16973.368 & 16876 & 63(4) & 16902.443 & 63.8(3)\\
$3s \ ^2S_{1/2} \rightarrow 4s \ ^2S_{1/2}$ & 25739.991 & &   & 25668.451 & 44.6(2) \\
$3s \ ^2S_{1/2} \rightarrow 3d \ ^2D_{5/2}$ & 29172.839 & &   & 29075.611 & 58.3(2) \\
$3s \ ^2S_{1/2} \rightarrow 3d \ ^2D_{3/2}$ & 29172.889 & &   & 29075.658 & 58.3(2) \\
$3s \ ^2S_{1/2} \rightarrow 4p \ ^2P_{1/2}$ & 30266.990 & 30124 & 53(4) & 30196.536 & 54.1(1) \\
$3s \ ^2S_{1/2} \rightarrow 4p \ ^2P_{3/2}$ & 30272.580 & 30130 & 59(4) & 30202.914 & 60.1(2) \\
$3s \ ^2S_{1/2} \rightarrow 5s \ ^2S_{1/2}$ & 33200.675 & &   &    33637.053 & 49.3(7) \\
$3s \ ^2S_{1/2} \rightarrow 4d \ ^2D_{5/2}$ & 34548.731 & &   &    34563.654 & 58.3(1) \\
$3s \ ^2S_{1/2} \rightarrow 4d \ ^2D_{3/2}$ & 34548.766 & &   &    34563.694 & 58.3(1) \\
$3s \ ^2S_{1/2} \rightarrow 4f \ ^2F_{5/2}$ & 34586.920 & &   &    34498.066 & 58.2(2) \\
$3s \ ^2S_{1/2} \rightarrow 4f \ ^2F_{7/2}$ & 34586.920 & &   &    34498.073 & 58.2(2) \\
$3s \ ^2S_{1/2} \rightarrow 5p \ ^2P_{1/2}$ & 35040.380 & &   &    36415.269 & 52.5(2.1) \\
$3s \ ^2S_{1/2} \rightarrow 5p \ ^2P_{3/2}$ & 35042.850 & &   &    36227.460 & 60.5(2.0) \\
\hline
\hline
\end{tabular}
\end{center}
\label{tab3}
\end{table*}

\subsection{Results in Na}
Using our RCC method, we calculate first IPs of several low-lying states for
different values of $\alpha$ and they are given in Table \ref{tab2}. In Table
\ref{tab3}, we report the excitation energies and $q$ values for different
transitions with respect to the ground state using the above IP results. 

We have also compared our calculated excitation energies with their 
corresponding experimental results for the considered transitions in Table
\ref{tab3}. As seen, our results are of sub-one percent accurate for most of 
the transitions. Using the accuracy of the calculated excitation energies,
we have also estimated the accuracy of the $q$ values and given them in the 
parenthesis of the results. We have also compared our results with other 
available theoretical results in the same table for a few transitions reported 
by Berengut et al \cite{berengut1} using different many-body methods. As
can be noticed, our results are more accurate than the results reported
in \cite{berengut1}. 

 We also find that some of the $q$ values from the higher transitions are of 
same order of magnitudes with the low-lying transitions implying that the 
fine structure constant variation studies can also be carried out using these
transitions in Na. It is also possible to use them to construct the anchor
lines.

\begin{table*}[t]
\caption{FS (in $MHz/fm^2$) and SMS (in $GHz \ amu$) constants of many low-lying states in Na from different works.}
\begin{center}
\begin{tabular}{lcccccc}
\hline
\hline
 Atomic & \multicolumn{2}{c}{\underline{Safronova \cite{safronova}}} & \underline{Berengut \cite{berengut2}} & \multicolumn{3}{c}{\underline{This work}} \\
 state & $F$ & $Z$ & $Z$ & $F$ & $Z$ I & $Z$ II  \\
\hline
  &  &  \\
$3s \ ^2S_{1/2}$ & $-36.825$ & 53.94 & 69 & $-37.044$ & $71.4(-10.9)$ & $72.8(-10.9)$ \\
$3p \ ^2P_{1/2}$ & $1.597$ & $-43.36$  & $-40$ & $1.726$ & $-40.5(-0.4)$ & $-41.3(-0.4)$  \\
$3p \ ^2P_{3/2}$ & $1.603$ & $-43.39$  & $-39$ & $1.765$ & $-39.3(-3.3)$ & $-39.2(-3.4)$ \\ 
$4s \ ^2S_{1/2}$ &   &   &   &  $-8.555$ & $20.7(-2.7)$ & $21.0(-2.7)$  \\
$3d \ ^2D_{5/2}$ & $-0.062$ & $-2.99$ &   & $-0.051$ & $-2.4(-0.4)$ & $-2.4(-0.4)$  \\
$3d \ ^2D_{3/2}$ & $-0.062$ & $-2.95$  &   & $-0.049$ & $-2.3(-0.2)$ & $-2.3(-0.2)$  \\
$4p \ ^2P_{1/2}$ &   &   &    & $0.582$ & $-14.5(-0.3)$ & $-14.7(-0.3)$ \\
$4p \ ^2P_{3/2}$ &   &   &    & $0.596$ & $-14.1(-1.2)$ & $-14.1(-1.2)$ \\
$5s \ ^2S_{1/2}$ &   &   &    & $-5.546$ & $14.4(-2.2)$ & $14.6(-2.2)$  \\
$4d \ ^2D_{5/2}$ &   &   &    & $-0.028$ & $-1.9(-0.2)$ & $-1.9(-0.2)$  \\
$4d \ ^2D_{3/2}$ &   &   &    & $-0.027$ & $-1.9(-0.1)$ & $-1.9(-0.1)$  \\
$4f \ ^2F_{5/2}$ &   &   &    & $-0.004$ & $0.2(~ 0)$ & $0.2( ~0)$  \\
$4f \ ^2F_{7/2}$ &   &   &    & $-0.004$ & $0.2(~ 0)$ & $0.2(~ 0)$  \\
$5p \ ^2P_{1/2}$ &   &   &    & $0.791$ & $-19.8(-0.5)$ & $-20.2(-0.6)$ \\
$5p \ ^2P_{3/2}$ &   &   &    & $0.755$ & $-17.9(-1.5)$ & $-17.9(-1.5)$ \\
\hline
\hline
\end{tabular}
\end{center}
\label{tab4}
\end{table*}
We have calculated $F$ and $Z$ using the same wavefunctions for different 
states those were used to evaluate the above IPs of the corresponding states.
We have reported these results in Table \ref{tab4}. As discussed in Sec. II,
it is possible to use the radial integrals given either by Eq. (\ref{eqn7}) 
or by Eq. (\ref{eqn9}) to evaluate the SMS constants. To verify the numerical
accuracy in the SMS constant calculations, we have used both the expressions
to evaluate them. We report results as $Z$ I when we have used the expression
given by Eq. (\ref{eqn7}) and as $Z$ II when we have used the expression given 
by Eq. (\ref{eqn9}). Excellent agreement between the results obtained using 
both the expressions indicate high numerical accuracy in the calculations. 

\begin{table*}[h]
\caption{Determination of ISs in $MHz$ for Na. Results given as "Reco" are the 
recommended values after the numerical inaccuracies and neglected correlation
effects in the calculations.}
\begin{center}
\begin{tabular}{lcccccc}
\hline
\hline
 Transition & \multicolumn{3}{c}{22-23} & \multicolumn{3}{c}{23-24} \\
  &  Z I & Z II & Reco & Z I & Z II & Reco \\
\hline
  &  &  \\
$3s \ ^2S_{1/2} \rightarrow 3p \ ^2P_{1/2}$ & $758.9$ & $763.3$ & $759(15)$ & $699.4$ & $703.4$ & $699(15)$ \\
$3s \ ^2S_{1/2} \rightarrow 3p \ ^2P_{3/2}$ & $757.2$ & $759.7$ & $757(15)$ & $697.7$ & $700.1$ & $698(15)$ \\
$3s \ ^2S_{1/2} \rightarrow 4s \ ^2S_{1/2}$ & $924.9$ & $927.1$ & $925(15)$ & $852.6$ & $854.6$ & $853(15)$ \\
$3s \ ^2S_{1/2} \rightarrow 3d \ ^2D_{5/2}$ & $1079.0$ & $1081.7$ & $1079(20)$ & $994.5$ & $997.1$ & $995(20)$ \\
$3s \ ^2S_{1/2} \rightarrow 3d \ ^2D_{3/2}$ & $1078.8$ & $1081.5$ & $1079(20)$ & $994.3$ & $996.9$ & $994(20)$  \\
$3s \ ^2S_{1/2} \rightarrow 4p \ ^2P_{1/2}$ & $1138.9$ & $1142.1$ & $1140(20)$ & $1049.8$ & $1052.7$ & $1050(20)$ \\
$3s \ ^2S_{1/2} \rightarrow 4p \ ^2P_{3/2}$ & $1138.4$ & $1141.1$ & $1140(20)$ & $1049.3$ & $1051.8$ & $1049(20)$ \\
$3s \ ^2S_{1/2} \rightarrow 5s \ ^2S_{1/2}$ & $1194.7$ & $1197.1$ & $1195(15)$ & $1101.3$ & $1103.5$ & $1101(15)$ \\
$3s \ ^2S_{1/2} \rightarrow 4d \ ^2D_{5/2}$ & $1255.7$ & $1258.4$ & $1256(20)$ & $1157.4$ & $1159.9$ & $1157(20)$ \\
$3s \ ^2S_{1/2} \rightarrow 4d \ ^2D_{3/2}$ & $1255.7$ & $1258.4$ & $1256(20)$ & $1157.4$ & $1159.9$ & $1157(20)$ \\
$3s \ ^2S_{1/2} \rightarrow 4f \ ^2F_{5/2}$ & $1249.4$ & $1252.2$ & $1249(15)$ & $1151.7$ & $1154.2$ & $1152(15)$ \\
$3s \ ^2S_{1/2} \rightarrow 4f \ ^2F_{7/2}$ & $1249.4$ & $1252.2$ & $1249(15)$ & $1151.7$ & $1154.2$ & $1152(15)$ \\
$3s \ ^2S_{1/2} \rightarrow 5p \ ^2P_{1/2}$ & $1350.7$ & $1354.2$ & $1351(20)$ & $1245.0$ & $1248.3$ & $1245(20)$ \\
$3s \ ^2S_{1/2} \rightarrow 5p \ ^2P_{3/2}$ & $1340.8$ & $1343.6$ & $1341(20)$ & $1236.0$ & $1238.5$ & $1236(20)$ \\
\hline
\hline
\end{tabular}
\end{center}
\label{tab5}
\end{table*}
We have also given the available results for the same properties from the other
works which were calculated using different relativistic many-body methods.
Safronova and Johnson \cite{safronova} have used two orders of Coulomb 
interaction in their MBPT(3) approach to evaluate these quantities whereas
Berengut \cite{berengut2} have used
a chain of diagrams through the Green's function technique with the 
Brueckner orbitals. The important difference between these two methods is 
that Safronova and Johnson have evaluated the above properties from different
correlation diagrams separately within MBPT(3) method whereas Berengut {\it 
et al} have considered both the FS and SMS interaction operators in the 
atomic Hamiltonian with a varying parameter and solved them self-consistently.
By plotting various energies for different values of the parameters, they
have extracted the IS constants. However, Berengut et al have scaled their
wavefunctions to obtain the excitation energies matching with the experimental 
results and the same wavefunctions are used to estimate the above results.
Obviously, this cannot explain the strength of the many-body method. In the
calculations by Berengut et al, the contributions only from the large 
components (only the first term of the radial expression given by Eq. 
(\ref{eqn7})) has been taken to calculate the SMS constants. Nevertheless, 
there is a large differences in the SMS results reported in 
\cite{safronova} and \cite{berengut2}. The explanation given by Berengut et al
for the possible reason of discrepancy between the two calculations 
that the MBPT(3) method may have poor convergence for the considered 
properties. In that case, it will be interesting to see how the all order 
calculations in the RCC approach match with their results.

It is seen from Table \ref{tab4} that our results for FS constants differ
only little bit from the MBPT(3) results of Safronova and Johnson. However,
there is a large difference in the SMS constant results between these works.
Safronova and Johnson have given explictly contributions from various 
correlation diagram within their MBPT(3) method and we discuss below
explcitly the contributions from various RCC diagrams to both FS and SMS
constants from where a comparison between the results at both the level
approximations can be made. A careful analysis shows that the CCSD(T) method
misses out some of the important triple excitation contributions to the SMS
calculations through its two-body terms ($Z_2$) that contribute through
the MBPT(3) method. To estimate these contributions, we
have used the perturbed triple excitation RCC operator given by Eq. 
(\ref{eqn22}) in Eq. (\ref{eqn23}) and evaluated their contributions. 
These contributions are given in the parenthesis of our results for $Z$.
As seen it seems though the large discrepancies between the MBPT(3) results
and the CCSD(T) results are due to these contributions. It is possible that
these extra triple excitation contributions may cancel with some of the
quadrupole excitations, so we have not included these perturbative contributions
 in our final results instead consider them as possible source of errors
for the determination of ISs as given below. Our SMS constant results match
reasonably well with the results those are reported by Berengut et al, but they
do not seem to agree each other after the inclusion of contributions from
the triple excitations.

\begin{table}[h]
\caption{Calculated IPs (in $cm^{-1}$) for different states in Mg$^+$ with different values of $\alpha$.}
\begin{center}
\begin{tabular}{lrrr}
\hline
\hline
 Atomic & IP with & IP with & IP with \\
 state & $\omega(0.0)$ & $\omega(+0.05)$ & $\omega(-0.05)$ \\
\hline
  &  &  \\
$3s \ ^2S_{1/2}$ & 121160.860 & 121170.521 & 121151.201 \\
$3p \ ^2P_{1/2}$ &  85538.955 &  85542.670 &  85535.245 \\
$3p \ ^2P_{3/2}$ &  85441.378 &  85440.205 &  85442.549 \\
$4s \ ^2S_{1/2}$ &  51418.868 &  51421.576 &  51416.161 \\
$3d \ ^2D_{5/2}$ &  49736.923 &  49736.855 &  49736.992 \\
$3d \ ^2D_{3/2}$ &  49736.186 &  49736.081 &  49736.290 \\
$4p \ ^2P_{1/2}$ &  40575.576 &  40576.861 &  40574.294 \\
$4p \ ^2P_{3/2}$ &  40550.112 &  40549.742 &  40550.481 \\
$5s \ ^2S_{1/2}$ &  28310.051 &  28311.277 &  28308.826 \\
$4d \ ^2D_{5/2}$ &  27845.866 &  27845.832 &  27845.902 \\
$4d \ ^2D_{3/2}$ &  27845.425 &  27845.368 &  27845.482 \\
$4f \ ^2F_{5/2}$ &  27433.775 &  27433.785 &  27433.766 \\
$4f \ ^2F_{7/2}$ &  27433.684 &  27433.687 &  27433.680 \\
$5p \ ^2P_{1/2}$ &  23523.731 &  23524.392 &  23523.071 \\
$5p \ ^2P_{3/2}$ &  23551.625 &  23551.443 &  23551.807 \\
\hline
\hline
\end{tabular}
\end{center}
\label{tab6}
\end{table}
Substituting our calculated FS and SMS constants given in Table \ref{tab4} in
Eq. \ref{eqn14a} and using the experimental excitation energies given in Table 
\ref{tab3}, we have calculated ISs for different transitions from 
the ground state between $^{22}$Na$- ^{23}$Na and $^{23}$Na$- ^{24}$Na.
These results are given in Table \ref{tab5}. We have calculated ISs using the
SMS constants given by both $Z$ I and $Z$ II. Since the main source of errors
in the IS calculations come from the SMS part, we have neglected the errors
in FS and we present the recommended values (given as Reco in the above table) 
considering errors both from the numerical calculations and neglected 
correlation effects. There are three experimental measurements of ISs are 
available only for $^{22}$Na$- ^{23}$Na in the $3s \ ^2S \rightarrow 3p \ ^2P_{1/2}$ and $3s \ ^2S \rightarrow 3p \ ^2P_{3/2}$ transitions
\cite{pescht, huber, gangrsky}. The experimental values of IS in the 
$3s \ ^2S \rightarrow 3p \ ^2P_{1/2}$ transition are 758.5(7) MHz \cite{pescht}
and 756.9(1.9) MHz \cite{huber} and our result for this transition is well
in agreement within its error bar. The experimental value of IS in the
$3s \ ^2S \rightarrow 3p \ ^2P_{3/2}$ transition is 757.72(24) MHz
\cite{gangrsky} and our result is also in agreement with it. Rcenetly,
Korol and Kozlov have employed a hybrid many-body approach by combining the
configuration interaction (CI) method with MBPT to calculate SMS constants and 
obtain IS as 775.8 MHz and 776.5 MHz for the $3s \ ^2S \rightarrow 3p \ ^2P_{1/2}$ and $3s \ ^2S \rightarrow 3p \ ^2P_{3/2}$ transitions after neglecting the
FS constributions \cite{korol}. Nevertheless, our results match with all
the studies and therefore our reported IS results in the other transitions
can also be assumed to give correct predictions within their error bars.
These results, indeed, will be useful for the experimentalists to carry out 
the IS measurements in these transitions. In fact, our SMS and FS constants
can also be used to derive IS for other intercombination transitions.
 
\begin{table*}[h]
\caption{Excitation energies and $q$ values in $cm^{-1}$ for different states of Mg$^+$.}
\begin{center}
\begin{tabular}{lcccc}
\hline
\hline
 Transition & Experiment & Others \cite{dzuba2} & \multicolumn{2}{c}{\underline{This work}} \\
  &  EE \cite{nist} & $q$ & EE & $q$ \\
\hline
  &  &  \\
$3s \ ^2S_{1/2} \rightarrow 3p \ ^2P_{1/2}$ & 35669.31 & 120 & 35621.905 & 119.0(2) \\
$3s \ ^2S_{1/2} \rightarrow 3p \ ^2P_{3/2}$ & 35760.88 & 211 & 35719.482 & 216.6(3) \\
$3s \ ^2S_{1/2} \rightarrow 4s \ ^2S_{1/2}$ & 69804.95 &  & 69741.992 & 139.1(1) \\
$3s \ ^2S_{1/2} \rightarrow 3d \ ^2D_{5/2}$ & 71490.19 &  & 71423.937 & 194.6(2) \\
$3s \ ^2S_{1/2} \rightarrow 3d \ ^2D_{3/2}$ & 71491.06 &  & 71424.674 & 195.3(2) \\
$3s \ ^2S_{1/2} \rightarrow 4p \ ^2P_{1/2}$ & 80619.50 &  & 80585.284 & 167.5(1) \\
$3s \ ^2S_{1/2} \rightarrow 4p \ ^2P_{3/2}$ & 80650.02 &  & 80610.748 & 200.6(1) \\
$3s \ ^2S_{1/2} \rightarrow 5s \ ^2S_{1/2}$ & 92790.51 &  & 92850.809 & 168.7(1) \\
$3s \ ^2S_{1/2} \rightarrow 4d \ ^2D_{5/2}$ & 93310.59 &  & 93314.994 & 193.90(1) \\
$3s \ ^2S_{1/2} \rightarrow 4d \ ^2D_{3/2}$ & 93311.11 &  & 93315.435 & 194.34(1) \\
$3s \ ^2S_{1/2} \rightarrow 4f \ ^2F_{5/2}$ & 93799.63 &  & 93727.085 & 193.0(2) \\
$3s \ ^2S_{1/2} \rightarrow 4f \ ^2F_{7/2}$ & 93799.75 &  & 93727.176 & 193.1(2) \\
$3s \ ^2S_{1/2} \rightarrow 5p \ ^2P_{1/2}$ & 97455.12 &  & 97637.129 & 180.0(4) \\
$3s \ ^2S_{1/2} \rightarrow 5p \ ^2P_{3/2}$ & 97468.92 &  & 97609.235 & 196.8(3) \\
\hline
\hline
\end{tabular}
\end{center}
\label{tab7}
\end{table*}

\subsection{Results in Mg$^+$}
In Table \ref{tab5}, we present the IPs of different low-lying states for 
Mg$^+$ with different values of $\alpha$. The trend of the IPs are same 
with Na states when $\alpha$ changes from its laboratory value. From the IP
results for $\alpha=\alpha_0$, we determine the calculated excitation energies
for various transitions with respect to the ground state and have given
them in Table \ref{tab7} along with their experimental results. We have also
estimated $q$ values from the IP results obtained from different $\alpha$. 

\begin{table*}[h]
\caption{FS (in $MHz/fm^2$) and SMS (in $GHz \ amu$) constants of many low-lying states in Mg$^+$ from different works.}
\begin{center}
\begin{tabular}{ccccccc}
\hline
\hline
 Atomic & \multicolumn{2}{c}{\underline{Safronova}} & \underline{Berengut \cite{berengut2}} & \multicolumn{3}{c}{\underline{This work}} \\
 state & $F$ & $Z$ & $Z$ & $F$ & $Z$ I & $Z$ II  \\
\hline
  &  &  \\
$3s \ ^2S_{1/2}$ & $-116.01$ & 38 & 83 & $-116.150$ & $73.8(-18.7)$ & $77.7(-18.9)$ \\
$3p \ ^2P_{1/2}$ & $9.800$  & $-324$ &  $-296$ & $10.126$ & $-315.8(-2.8)$ & $-320.7(-2.8)$ \\
$3p \ ^2P_{3/2}$ & $9.811$  & $-323$ &  $-290$ & $10.226$ & $-311.7(-10.2)$ & $-311.7(-10.3)$\\ 
$4s \ ^2S_{1/2}$ &   &   &   & $-31.833$ & $45.1(-5.3)$ & $46.1(-5.3)$  \\
$3d \ ^2D_{5/2}$ & $-0.085$  & $-106$ &   &  $-0.016$ & $-100.5(-5.6)$ & $-100.4(-5.6)$ \\
$3d \ ^2D_{3/2}$ & $-0.083$  & $-105$ &    & $-0.006$ & $-99.6(-2.4)$ & $-99.5(-2.4)$ \\
$4p \ ^2P_{1/2}$ &  &   &   & $3.470$ & $-110.5(-1.8)$ & $-112.2(-1.8)$ \\
$4p \ ^2P_{3/2}$ &  &   &   & $3.498$ & $-109.2(-3.8)$ & $-109.2(-3.9)$ \\
$5s \ ^2S_{1/2}$ &  &   &   & $-14.283$ & $23.3(-2.4)$ & $23.7(-2.4)$  \\
$4d \ ^2D_{5/2}$ &  &   &   & $0.035$ & $-54.2(-3.1)$ & $-54.2(-3.1)$ \\
$4d \ ^2D_{3/2}$ &  &   &   & $0.039$ & $-53.8(-1.4)$ & $-53.7(-1.4)$  \\
$4f \ ^2F_{5/2}$ &  &   &   & $-0.021$ & $0.7(-0.1)$ & $0.7(-0.1)$  \\
$4f \ ^2F_{7/2}$ &  &   &   & $-0.021$ & $0.7(-0.1)$ & $0.7(-0.1)$  \\
$5p \ ^2P_{1/2}$ &  &   &   & $1.777$ & $-56.8(-1.1)$ & $-57.7(-1.1)$ \\
$5p \ ^2P_{3/2}$ &  &   &   & $1.786$ & $-55.9(-2.0)$ & $-55.9(-2.1)$ \\
\hline
\hline
\end{tabular}
\end{center}
\label{tab8}
\end{table*}
As seen from the above table, our excitation energies for the given transitions
match well with the experimental results indicating that our calculations 
are very accurate. Using these accuracies, we have estimated the errors associated with 
the determined $q$ values and they are given in the parenthesis in the same
table. There are also calculations of $q$ values available for the first two transitions which are
given in the above table and unlike the results for Na, these results differ 
slightly from ours. By comparing Table \ref{tab3} and Table \ref{tab7}, it
is clear that the $q$ values of the given transitions are larger in Mg$^+$ 
suggesting this candidate is more suitable than Na to carry out the fine
structure variation study. Since both the laser cooling and ion trapping 
techniques are well advanced these days, very precise $\alpha$ variation 
measurement can be persued in Mg$^+$. Preliminary works by Batteiger et al
\cite{batteiger} on fine structure and IS measurements are the initial steps 
and recent motivation for the theorists towards such studies.

\begin{table*}[h]
\caption{Determination of ISs in $MHz$ for Mg$^+$. Results given as "Reco" are 
the recommended values after the numerical inaccuracies and neglected 
correlation effects in the calculations.}
\begin{center}
\begin{tabular}{lcccccc}
\hline
\hline
 Transition & \multicolumn{3}{c}{24-25} & \multicolumn{3}{c}{24-26} \\
  &  Z I & Z II & Reco & Z I & Z II & Reco \\
\hline
  &  &  \\
$3s \ ^2S_{1/2} \rightarrow 3p \ ^2P_{1/2}$ & $1603.9$ & $1618.6$ & $1604(25)$ & $3077.0$ & $3105.2$ & $3077(40)$ \\
$3s \ ^2S_{1/2} \rightarrow 3p \ ^2P_{3/2}$ & $1599.7$ & $1606.2$ & $1600(40)$ & $3069.0$ & $3081.5$ & $3069(60)$ \\
$3s \ ^2S_{1/2} \rightarrow 4s \ ^2S_{1/2}$ & $1946.6$ & $1951.5$ & $1947(25)$ & $3735.3$ & $3744.6$ & $3735(40)$ \\
$3s \ ^2S_{1/2} \rightarrow 3d \ ^2D_{5/2}$ & $2229.7$ & $2236.0$ & $2230(25)$ & $4278.2$ & $4290.4$ & $4278(40)$ \\
$3s \ ^2S_{1/2} \rightarrow 3d \ ^2D_{3/2}$ & $2228.2$ & $2234.5$ & $2228(25)$ & $4275.4$ & $4287.5$ & $4275(40)$  \\
$3s \ ^2S_{1/2} \rightarrow 4p \ ^2P_{1/2}$ & $2497.3$ & $2506.6$ & $2497(30)$ & $4791.8$ & $4809.8$ & $4792(50)$ \\
$3s \ ^2S_{1/2} \rightarrow 4p \ ^2P_{3/2}$ & $2495.8$ & $2502.3$ & $2496(30)$ & $4789.0$ & $4801.5$ & $4789(50)$ \\
$3s \ ^2S_{1/2} \rightarrow 5s \ ^2S_{1/2}$ & $2614.2$ & $2620.1$ & $2614(30)$ & $5016.5$ & $5027.7$ & $5017(50)$ \\
$3s \ ^2S_{1/2} \rightarrow 4d \ ^2D_{5/2}$ & $2753.6$ & $2760.1$ & $2754(30)$ & $5283.8$ & $5296.3$ & $5284(50)$ \\
$3s \ ^2S_{1/2} \rightarrow 4d \ ^2D_{3/2}$ & $2752.9$ & $2759.3$ & $2753(30)$ & $5282.5$ & $5294.7$ & $5283(50)$ \\
$3s \ ^2S_{1/2} \rightarrow 4f \ ^2F_{5/2}$ & $2673.2$ & $2679.7$ & $2673(30)$ & $5129.5$ & $5142.0$ & $5130(50)$ \\
$3s \ ^2S_{1/2} \rightarrow 4f \ ^2F_{7/2}$ & $2673.2$ & $2679.7$ & $2673(30)$ & $5129.5$ & $5142.0$ & $5130(50)$ \\
$3s \ ^2S_{1/2} \rightarrow 5p \ ^2P_{1/2}$ & $2876.3$ & $2884.3$ & $2876(30)$ & $5519.3$ & $5534.6$ & $5519(50)$ \\
$3s \ ^2S_{1/2} \rightarrow 5p \ ^2P_{3/2}$ & $2874.0$ & $2880.5$ & $2874(30)$ & $5514.9$ & $5527.4$ & $5515(50)$ \\
\hline
\hline
\end{tabular}
\end{center}
\label{tab9}
\end{table*}
We have calculated the IS parameters in Mg$^+$ using the same wavefunctions 
those are used to calculate IPs of the corresponding states. They are presented
in Table \ref{tab8}. Again, we have determined SMS constants as $Z$ I and $Z$ 
II using different radial expressions as it has been discussed before. There
is a large difference between both the results in Mg$^+$ where it was in
excellent agreement for Na. This clearly indicates that the wavefunctions
behave differently in both Na and Mg$^+$ although similar basis functions are
used in both the systems except their different nuclear sizes.

We have also compared FS and SMS constants obtained from other works with ours
in the above table. Unlike the case for Na, our results do not match well
with the results obtained by Berengut et al \cite{berengut2}. The difference 
between their works and ours is already discussed in the previous subsection.
It seems though our $3p$ results are close to the results obtained by Safronova
and Johnson \cite{safronova}, but the SMS constant result for the ground state 
do not match at all from these two works. In fact after inclusion of the triple
excitation effects perturbatively, as discussed in the previous subsection, the
results do not seem to agree with any of the above works. In our calculations,
these contributions from the perturbed triple excitation effects are 
considered as a possibel source of errors in the calculations of SMS constants.
The FS constant results between Safronova and Johnson and ours match reasonably
except for the $3d$ states. Contributions from various correlation effects 
in the FS and SMS constant calculations through various RCC terms for this
system are given explicitly below.

\begin{table*}[h]
\caption{Comparative contributions from different RCC terms to the FS constants in Na and Mg$^+$. $\overline{F}$ is the effective one-body terms of 
$e^{T^{\dagger}}Fe^T$. Contributions from the effective two-body and three-body terms of above non-truncative exponetials are given as "Others" along with the
small contributions from normalization of the wavefunctions.}
\begin{center}
\begin{tabular}{l cc cc cc cc cc}
\hline
\hline
 Atomic & \multicolumn{2}{c}{\underline{DF(F)}} & \multicolumn{2}{c}{\underline{$ \overline{F}$}} & \multicolumn{2}{c}{\underline{$ \overline{F}S_{1v}+c.c.$}} & \multicolumn{2}{c}{\underline{$\overline{F}S_{2v}+c.c.$}} & \multicolumn{2}{c}{\underline{Others}} \\
 state & Na & Mg$^+$ & Na & Mg$^+$ & Na & Mg$^+$ & Na & Mg$^+$ & Na & Mg$^+$ \\
\hline
  &  &  \\
$3s \ ^2S_{1/2}$ & $-29.749$ & $-104.686$ & $-30.259$ & $-105.838$ & $-4.724$ & $-9.264$ & $-1.700$ & $-0.693$ & $-0.361$ & $-0.355$ \\
$3p \ ^2P_{1/2}$ & $-0.008$ & $-0.059$ & 0.032 & 0.042 & 0.003 & $3\times10^{-4}$ & $1.566$ & 9.590 & $0.125$ & 0.495 \\
$3p \ ^2P_{3/2}$ & $\sim 0$ & $\sim 0$ & 0.055 & 0.151 & 0.006 & 0.010 & $1.562$ & 9.537 & $0.143$ & 0.528 \\
$4s \ ^2S_{1/2}$ & $-7.239$ & $-29.666$ & $-7.347$ & $-29.912$ & $-0.818$ & $-1.892$  & $-0.326$ & 0.088 & $-0.074$ & $-0.117$ \\
$3d \ ^2D_{5/2}$ & $\sim 0$ & $\sim 0$  & $7\times10^{-5}$  & 0.002 & $1\times10^{-5}$  & $2\times10^{-4}$  & $-0.047$ & $-0.008$ & $-0.004$ & $-0.001$ \\
$3d \ ^2D_{3/2}$ & $\sim 0$  & $\sim 0$ & $2\times10^{-5}$ & $3\times10^{-4}$ & $2\times10^{-6}$ & $2\times10^{-5}$ & $-0.047$ & $-0.006$ & $-0.002$ & $2\times10^{-4}$ \\
$4p \ ^2P_{1/2}$ & $-0.003$ & $-0.021$ & 0.010 & 0.009 & 0.001 & $-0.001$ & $0.534$ & 3.316 & $0.037$ & 0.145 \\
$4p \ ^2P_{3/2}$ & $\sim 0$ & $\sim 0$ & 0.017 & 0.045 & 0.002  & 0.002 & $0.534$ & 3.297 & $0.043$ & 0.154 \\
$5s \ ^2S_{1/2}$ & $-5.045$  & $-13.539$ & $-5.117$ & $-13.641$ & $-0.222$ & $-0.670$ & $-0.193$ & $0.079$ & $-0.014$ & $-0.051$ \\
$4d \ ^2D_{5/2}$ & $\sim 0$  & $\sim 0$  & $5\times10^{-5}$  & 0.001  & $6\times10^{-6}$  & $1\times10^{-4}$ & $-0.026$ & 0.035 & $-0.002$ & $-0.001$ \\
$4d \ ^2D_{3/2}$ & $\sim 0$ & $\sim 0$  & $1\times10^{-5}$  & $1\times10^{-4}$  & $1\times10^{-6}$  & $1\times10^{-5}$  & $-0.026$ & 0.036  & $-0.001$ & 0.003\\
$4f \ ^2F_{5/2}$ & $\sim 0$ & $\sim 0$ & $\sim 0$  & $3\times10^{-6}$ & $\sim 0$  & $\sim 0$  & $-0.003$ & $-0.021$ & $4\times10^{-4}$ & $-9\times10^{-5}$ \\
$4f \ ^2F_{7/2}$ & $\sim 0$  & $\sim 0$  & $\sim 0$  & $6\times10^{-6}$  & $\sim 0$  & $\sim 0$  & $-0.003$ & $-0.021$ & $4\times10^{-4}$ & $-9\times10^{-5}$ \\
$5p \ ^2P_{1/2}$ & $-0.004$ & $-0.011$  & 0.013 & 0.004 & $2\times10^{-4}$ & $-4\times10^{-4}$ & 0.748 & 1.706 & $0.030$ & 0.066 \\
$5p \ ^2P_{3/2}$ & $\sim 0$ & $\sim 0$ & 0.022 & 0.022  & 0.001 & 0.001 & $0.697$ & 1.693 & $0.035$ & 0.070 \\
\hline
\hline
\end{tabular}
\end{center}
\label{tab10}
\end{table*}
Using the above FS, SMS results and excitations energies given in Table 
\ref{tab7}, we have determined the ISs in various transitions with respect to 
the ground state between $^{24}$Mg$^+- ^{25}$Mg$^+$ and $^{24}$Mg$^+- ^{26}$Mg$^+$. These results are given in Table \ref{tab9}. There are experimental 
measurements of ISs are available only in the $3s \ ^2S \rightarrow 3p \ ^2P_{1/2}$ and 
$3s \ ^2S \rightarrow 3p \ ^2P_{3/2}$ transitions between $^{24}$Mg$^+- ^{26}$Mg$^+$ \cite{drullinger, batteiger} among which Batteiger et al \cite{batteiger}
results are more recent and precise. The experimental values of IS in the
$3s \ ^2S \rightarrow 3p \ ^2P_{3/2}$ transition are 3050(100) MHz
\cite{drullinger} and 3087.560(87) MHz \cite{batteiger} and our result is
in good agreement with them. The experimental values of IS in the $3s \ ^2S \rightarrow 3p \ ^2P_{3/2}$ transition are 3050(100) MHz \cite{drullinger} and 
3087.560(87) MHz \cite{batteiger} and our result is also in agreement with them.
Korol and Kozlov have obtained this result as 3086.3 MHz using the CI$+$MBPT 
approach and neglecting the FS contribution \cite{korol}. Agreement
between all these results shows that our IS results reported for the other
transitions will also be reasonable accurate and these results will guide 
any new experiments to measure IS in any of these transitions in the right
direction. Using our FS and SMS constants from various states, it is also
possible to determine IS in many more intercombination lines.

\begin{figure}[h]
\includegraphics[width=8.5cm,clip=true]{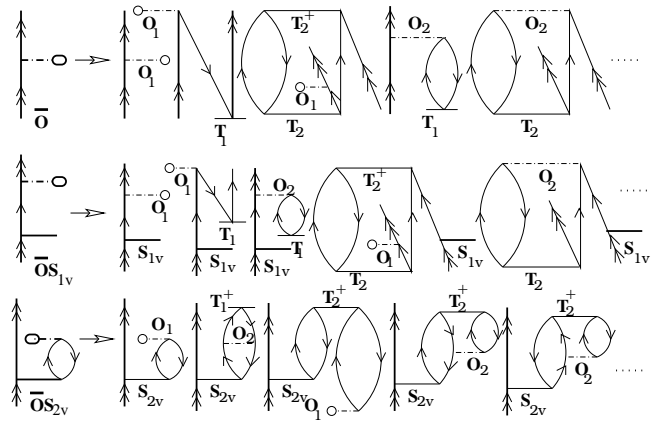}
\caption{Contributions from important RCC terms through the effective one-body
operators.}
\label{fig4}
\end{figure}

\subsection{Comparative studies of correlation effects}
In Table \ref{tab10}, we have reported contributions from DF and various RCC
terms for the FS constant calculations of all the reported states both in
Na and Mg$^+$. This
trend can be matched with the individual contributions reported by Safronova 
and Johnson \cite{safronova} to find out how the all order correlation effects 
differ compared to MBPT(3). We have shown diagrammatically from the important
contributing effective one-body RCC terms in Fig. \ref{fig4} and expand the effective operators
into a series of diagrams that contribute altogether through these terms (the diagrams
coming from the two-body operator form of $O$ do not contribute here). It is 
obvious from this figure that the difference between $\overline{F}$ and $F$ 
correspond to the core correlation effects which seem to be small for this 
property in both the systems. $\overline{F}S_{1v}$ and its conjugate terms
account all order correlation effects from the pair-correlation type diagrams.
Further, $\overline{F}S_{2v}$ and its conjugate terms consider all order
correlation effects from the core-polarization diagrams. From the above table,
we find that the pair correlation effects are stronger in the $s$ states
whereas the core-polarization effects play crucial role in obtaining the final
results in all other states for FS constant calculations. Contributions from
non-linear terms and normalization of the wavefunctions are given as "Others"
and they seem to be small.

\begin{figure}[h]
\includegraphics[width=8.5cm,clip=true]{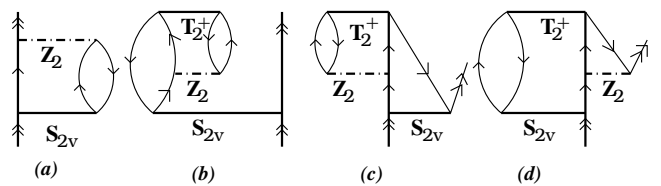}
\caption{Contributions from important effective two-body RCC terms to the specific mass shift calculations.}
\label{fig5}
\end{figure}

 We present in Table \ref{tab11} the individual contributions to both $Z$ I 
and $Z$ II from various RCC terms for various states in Na. Along with the 
contributions from the terms given for FS calculations, we also give the
contributions from two-body term $Z_2 S_{2v}$ that contributes
significantly in this calculation. The corresponding diagrammatic 
representation of this term is shown in Fig. \ref{fig5}(a). This seems to
be a different types of pair correlation effects. As seen from the above
table, the core correlation effects are sizably bigger in this property
and the core-polarization effects coming through $\overline{Z_1}S_{2v}$
contribute larger than the pair correlation effects from 
$\overline{Z_1}S_{1v}$ with opposite sign. Contributions from $Z_2 S_{2v}$ 
are also larger than $\overline{Z_1}S_{1v}$ with opposite sign. As a
result the final results have different signs than the DF results for most
of states implying there are strong correlation effects in this property. 
In fact, there are also other
non-linear two-body terms contribute significantly and their contributions
are given by "Others" along with the contributions from normalization of the 
wavefunctions. Particularly the non-linear RCC terms of 
$T_2^{\dagger} Z_2 S_{2v}$ form contribute the most to "Others" and some
of such diagrams are shown in Fig. \ref{fig5}(b-d).

\begin{figure}[h]
\includegraphics[width=8.8cm,clip=true]{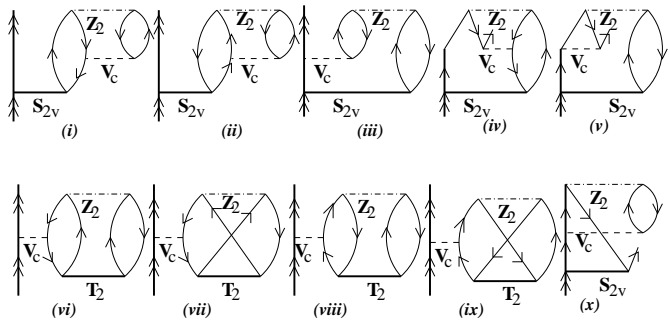}
\caption{Important triple excitation diagrams coming through $S_{3v}^{pert}$ perturbative operator after contracting with $Z_2$.}
\label{fig6}
\end{figure}
In Table \ref{tab12}, we present the contributions from different RCC terms,
that are discussed above, to the SMS constant calculations for Mg$^+$. There
seem to be little larger differences at the DF results for $Z$ I and $Z$ II
in this system. The trend of the results in this case is almost similar to
Na, but the magnitudes are bigger for all the states.

\begin{table*}[h]
\caption{Contributions from different RCC terms to $Z$ I and $Z$ II in Na. $\overline{Z}$ is again the effective one-body terms of $e^{T^{\dagger}}Ze^T$. 
Note that $Z=Z_1+Z_2$ here. Important contributions from $Z_2 S_{2v}$ and
its conjugate term are given explicitly. Other higher order contributions
from effective two-body, three-body etc operators of the above non-terminating
series with the normalization of the wavefunctions are given by "Others".}
\begin{center}
\begin{tabular}{l cc cc cc cc cc cc}
\hline
\hline
 Atomic & \multicolumn{2}{c}{\underline{DF($Z_1$)}} & \multicolumn{2}{c}{\underline{$ \overline{Z}$}} & \multicolumn{2}{c}{\underline{$ \overline{Z_1}S_{1v}+c.c.$}} & \multicolumn{2}{c}{\underline{$\overline{Z_1}S_{2v}+c.c.$}} & \multicolumn{2}{c}{\underline{$Z_2 S_{2v}+c.c.$}} &\multicolumn{2}{c}{\underline{Others}} \\
 state & $Z$ I & $Z$ II & $Z$ I & $Z$ II & $Z$ I & $Z$ II & $Z$ I & $Z$ II & $Z$ I & $Z$ II & $Z$ I & $Z$ II \\
\hline
  &  &  \\
$3s \ ^2S_{1/2}$ & $-221.999$ & $-221.633$ & $-196.613$ & $-196.195$ & $-23.627$ & $-23.571$ & $163.268$ & $164.040$ & 74.726 & 74.760 & $53.683$ & 53.785 \\
$3p \ ^2P_{1/2}$ & $-115.614$ & $-116.459$ & $-117.300$ & $-118.183$ & $-18.601$ & $-18.742$ & $51.330$ & $51.436$ & 26.707 & 26.779 & $17.404$ & 16.975 \\
$3p \ ^2P_{3/2}$ & $-115.519$  & $-115.740$  & $-115.762$ & $-115.988$ & $-18.366$  & $-18.404$ & $50.786$ & $51.054$ & $26.696$ & 26.756 & $17.349$ & 17.350 \\
$4s \ ^2S_{1/2}$ & $-49.485$ & $-49.400$ & $-43.613$ & $-43.516$ & $-3.130$ & $-3.122$ & $37.071$ & 37.256 & 18.319 & 18.327 & $12.083$ & 12.090 \\
$3d \ ^2D_{5/2}$ & $-4.879$  & $-4.880$ & $-5.816$ & $-5.817$ & $-0.705$ & $-0.705$ & $1.888$ & 1.895 & $-0.097$ & $-0.096$ & $2.354$ & 2.357 \\
$3d \ ^2D_{3/2}$ & $-4.845$  & $-4.846$ & $-5.771$ & $-5.773$ & $-0.700$ & $-0.701$ & $1.873$ & 1.883 & $-0.101$ & $-0.100$ & $2.352$ & 2.220 \\
$4p \ ^2P_{1/2}$ & $-38.960$ & $-39.244$ & $-39.443$ & $-39.739$ & $-5.510$ & $-5.552$ & $16.560$ & 16.594 & $8.827$ & 8.851 & 5.108 & 5.108 \\
$4p \ ^2P_{3/2}$ & $-39.021$ & $-39.096$ & $-39.041$ & $-39.118$ & $-5.460$  & $-5.472$ & $16.426$ & 16.512 & $8.847$ & 8.867 & 5.142 & 5.148 \\
$5s \ ^2S_{1/2}$ & $-33.580$  & $-33.523$ & $-29.557$  & $-29.492$ & $-0.137$ &  $-0.136$ & $24.679$ & 24.795 & $12.583$ & 12.588 & 6.827 & 6.836 \\
$4d \ ^2D_{5/2}$ & $-3.510$ & $-3.510$ & $-4.183$ & $-4.184$ & $-0.462$ & $-0.462$ & $1.169$ & 1.173 & $0.178$ & 0.178 & 1.365 & 1.368 \\
$4d \ ^2D_{3/2}$ & $-3.486$ & $-3.487$ & $-4.152$ & $-4.153$ & $-0.459$ & $-0.459$ & $1.158$ & 1.165 & $0.172$ & 0.173 & 1.356 & 1.357 \\
$4f \ ^2F_{5/2}$ & $0.0$ & 0.0 & $1\times10^{-6}$  & $1\times10^{-6}$ & $\sim 0$  & $\sim 0$  & $0.092$ & 0.092 & $-0.101$ & $-0.101$ & $0.182$ & 0.183 \\
$4f \ ^2F_{7/2}$ & $0.0$  & 0.0  & $3\times10^{-6}$ & $3\times10^{-6}$ & $\sim 0$  & $\sim 0$ & $0.091$ & 0.092 & $-0.101$ & $-0.101$ & $0.183$ & 0.183 \\
$5p \ ^2P_{1/2}$ & $-55.689$ & $-56.094$ & $-56.330$ & $-56.752$ & $-4.579$ & $-4.614$ & 22.698 & 22.744 & $12.890$ & 12.923 & $5.543$ & 4.991 \\
$5p \ ^2P_{3/2}$ & $-51.994$ & $-52.094$ & $-51.984$ & $-52.086$ & $-4.261$ & $-4.271$ & 21.006 & 21.116 & $12.046$ & 12.072 & $5.272$ & 5.276 \\
\hline
\hline
\end{tabular}
\end{center}
\label{tab11}
\end{table*}
In order to understand the role of triple excitation effects in the SMS 
constant calculations those do not appear at the CCSD(T) method, we have
given contributions from the important perturbative terms in Table 
\ref{tab13} whose diagrammatic representation are shown in Fig. \ref{fig6}.
It is obvious from this table that the individua contributions from triple
excitation effects are large, but the final contributions are small due to
large cancellation among themselves. In fact, we believe that these results will
 also be further cancelled with the contributions from quadrupole excitations
which was not possible to consider here due to their computational complexity.
Again, triple excitation effects in Mg$^+$ seem to be larger than Na as
per the magnitude of their final results. Contributions from triple excitations
to FS constant calculations are very small compared to the CCSD(T) results 
due to the fact that it is a scalar one-body operator and these contributions
have been neglected here.

\begin{table*}[h]
\caption{Contributions from different RCC terms to $Z$ I and $Z$ II in Mg$^+$. 
$\overline{Z}$ is as defined for Na. Important contributions from $Z_2 S_{2v}$ 
and its conjugate term are given explicitly. Other higher order contributions
from effective two-body, three-body etc operators of the above non-terminating
series with the normalization of the wavefunctions are given by "Others".}
\begin{center}
\begin{tabular}{l cc cc cc cc cc cc}
\hline
\hline
 Atomic & \multicolumn{2}{c}{\underline{DF($Z_1$)}} & \multicolumn{2}{c}{\underline{$ \overline{Z}$}} & \multicolumn{2}{c}{\underline{$ \overline{Z_1}S_{1v}+c.c.$}} & \multicolumn{2}{c}{\underline{$\overline{Z_1}S_{2v}+c.c.$}} & \multicolumn{2}{c}{\underline{$Z_2 S_{2v}+c.c.$}} &\multicolumn{2}{c}{\underline{Others}} \\
 state & $Z$ I & $Z$ II & $Z$ I & $Z$ II & $Z$ I & $Z$ II & $Z$ I & $Z$ II & $Z$ I & $Z$ II & $Z$ I & $Z$ II \\
\hline
  &  &  \\
$3s \ ^2S_{1/2}$ & $-563.474$ & $-562.198$ & $-491.841$ & $-490.453$ & $-31.853$ & $-31.751$ & $367.573$ & $369.704$ & 155.300 & 155.426 & $74.612$ & 74.740 \\
$3p \ ^2P_{1/2}$ & $-599.456$ & $-604.697$  & $-601.684$ & $-606.979$ & $-61.947$ & $-62.504$ & $197.041$ & 197.575 & 112.017 & 112.417 & $38.791$ & 38.787 \\
$3p \ ^2P_{3/2}$ & $-597.858$  & $-599.283$ & $-595.895$ & $-597.356$ & $-61.373$ & $-61.534$ & $194.989$ & 196.228 & $111.680$ & 112.015 & $38.929$ & 38.970 \\
$4s \ ^2S_{1/2}$ & $-137.727$ & $-137.395$ & $-119.996$ & $-119.632$ & $-4.302$ & $-4.286$  & $97.601$ & 98.165 & 51.151 & 51.183 & $20.605$ & 20.622 \\
$3d \ ^2D_{5/2}$ & $-120.967$ & $-120.988$ & $-140.895$ & $-140.932$ & $-14.887$ & $-14.892$ & $23.333$ & 23.420 & $16.100$ & 16.120 & $15.895$ & 15.916 \\
$3d \ ^2D_{3/2}$ & $-120.263$ & $-120.311$ & $-139.972$ & $-140.039$ & $-14.822$ & $-14.831$ & $23.108$ & 23.270 & $15.992$ & 16.013 & $16.098$ & 16.123 \\
$4p \ ^2P_{1/2}$ & $-210.097$ & $-211.898$  & $-210.681$ & $-212.527$ & $-18.068$ & $-18.231$ & $66.176$ & 66.359 & $40.079$ & 40.218 & 11.961 & 11.955 \\
$4p \ ^2P_{3/2}$ & $-209.467$ & $-209.969$ & $-208.762$ & $-209.278$ & $-17.899$ & $-17.947$ & $65.497$ & 65.910 & $39.939$ & 40.057 & 12.028 & 12.034 \\
$5s \ ^2S_{1/2}$ & $-60.074$  & $-59.927$ & $-52.339$ & $-52.178$ & $-1.102$ & $-1.097$ & $43.551$ & 43.802 & $24.170$ & 24.184 & 9.032 & 9.037 \\
$4d \ ^2D_{5/2}$ & $-66.429$ & $-66.441$ & $-77.396$ & $-77.418$ & $-7.602$ & $-7.604$ & $11.985$ & 12.028 & $11.232$ & 11.244 & 7.566 & 7.575 \\
$4d \ ^2D_{3/2}$ & $-66.071$ & $-66.099$ & $-76.921$ & $-76.960$ & $-7.575$ & $-7.579$ & $11.867$ & 11.951 & $11.169$ & 11.181 & 7.665 & 7.674 \\
$4f \ ^2F_{5/2}$ & $0.0$  & 0.0  & $9\times10^{-5}$  & $9\times10^{-5}$ & $4\times10^{-6}$  & $4\times10^{-6}$ & $0.749$ & 0.753 & $-1.465$ & $-1.465$ & $1.387$ & 1.389 \\
$4f \ ^2F_{7/2}$ & $0.0$  & 0.0  & $2\times10^{-4}$  & $2\times10^{-4}$ & $8\times10^{-6}$ & $8\times10^{-6}$  & $0.753$ & 0.757 & $-1.473$ & $-1.473$ & $1.384$ & 1.387 \\
$5p \ ^2P_{1/2}$ & $-108.722$ & $-109.654$ & $-109.041$ & $-109.996$ & $-8.459$ & $-8.535$ & 33.751 & 33.846 & $20.979$ & 21.051 & $5.987$ & 5.982 \\
$5p \ ^2P_{3/2}$ & $-108.340$ & $-108.600$  & $-107.987$ & $-108.254$ & $-8.149$ & $-8.171$  & 33.359 & 33.569 & $20.893$ & 20.953 & $5.933$ & 5.934 \\
\hline
\hline
\end{tabular}
\end{center}
\label{tab12}
\end{table*}

\begin{table*}[h]
\caption{Contributions from some of the important triple excitations comining
through the contraction of $S_{3v}^{pert}$ and $Z_2$ operators.}
\begin{center}
\begin{tabular}{l cc cc cc cc cc}
\hline
\hline
 Atomic & \multicolumn{2}{c}{\underline{$3s \ ^2S_{1/2}$}} & \multicolumn{2}{c}{\underline{$3p \ ^2P_{1/2}$}} & \multicolumn{2}{c}{\underline{$3p \ ^2P_{3/2}$}} & \multicolumn{2}{c}{\underline{$3d \ ^2D_{3/2}$}} & \multicolumn{2}{c}{\underline{$3d \ ^2D_{5/2}$ }} \\
 state & Na & Mg$^+$ & Na & Mg$^+$ & Na & Mg$^+$ & Na & Mg$^+$ & Na & Mg$^+$ \\
\hline
  &  &  \\
Diag. $(i)$ & $-3.596$ & $-9.191$ & $-1.795$ & $-7.057$ & $-1.795$ & $-7.052$ & $-0.050$ & $-0.438$ & $-0.050$ & $-0.436$ \\
Diag. $(ii)$ & $-10.960$ & $-15.098$ & $-4.553$ & $-10.008$ & $-4.553$ & $-9.998$ & $-0.251$ & $-1.606$ & $-0.251$ & $-1.604$ \\
Diag. $(iii)$ & 16.895 & 31.263 & 3.934 & 10.260 & 1.908 & 4.283 & 0.306 & 2.737 & 0.205 & 1.924 \\
Diag. $(iv)$ & $-3.896$ & $-9.512$ & $-1.565$ & $-5.952$ & $-1.588$ & $-6.034$ & $-1 \times 10^{-4}$ & $-0.008$ & $-1 \times 10^{-4}$ & $-0.008$ \\
Diag. $(v)$ & $-4.916$ & $-10.144$ & $-0.157$ & $-1.136$ & 0.609 & 1.792 & $-0.003$ & $-0.081$ & $-0.030$ & $-0.626$ \\
Diag. $(vi)$ & $650.914$ & 878.079 & 458.160 & 718.768 & 457.730 & 717.810 & 253.550 & 480.306 & 253.554 & 480.327 \\
Diag. $(vii)$ & $-137.896$ & $-180.893$ & $-96.981$ & $-147.981$ & $-96.890$ & $-147.783$ & $-53.611$ & $-98.642$ & $-53.611$ & $-98.646$ \\
Diag. $(viii)$ & $-646.518$ & $-868.182$ & $-456.165$ & $-711.433$ & $-455.733$ & $-710.482$ & $-253.546$ & $-480.185$ & $-253.549$ & $-480.206$ \\
Diag. $(ix)$ & 136.936 & 178.805 & 96.575 & 146.549 & 96.483 & 146.351 & 53.609 & 98.603 & 53.610 & 98.607 \\
Diag. $(x)$ & $-10.803$ & $-15.444$ & 0.150 & 0.004 & 0.765 & 1.455 & $-0.068$ & $-0.731$ & $-0.186$ & $-2.590$ \\
\hline
\hline
\end{tabular}
\end{center}
\label{tab13}
\end{table*}

\section{Summary}
In summary, we have employed the relativistic coupled-cluster theory to
estimate the fine structure constant variation coefficients for many
transitions accurately in sodium and singly ionized magnesium. We have also 
calculated isotope shifts for the corresponding transitions in the
considered systems. From these results, it is possible to construct 
suitable anchor and probe lines for the proposed fine structure constant 
variation studies in the above systems. We have also compared our 
results with other reported results obtained using different many-body approaches.
By giving contributions from individual terms, we have shown explicitly
the role of all order correlation effects in the isotope shift properties.
We have also tried to understand the reasons for the differences in the results obtained
from other calculations by finding contributions from important triple
excitations. Our calculated results those are reported for the first time 
would be the bench mark results for the experimentalists to carry out
their measurements in the right direction.

\section*{Acknowledgment}
The author is grateful to V. Batteiger, M. Kozlov, M. Kowalska and B. P. Das 
for useful discussions and thanks V. Athalye for her help during this work.

\end{document}